\documentstyle[12pt]{article}

\advance\textheight by \topskip

\newcommand{\doublespace}{
  \renewcommand{\baselinestretch}{1.75}
  \large\normalsize}
\newcommand{\td}{\tau_d}
\newcommand{\erinf}{E_\infty}
\newcommand{\eo}{E_i}
\newcommand{\ero}{E_0}
\newcommand{\eat}{\tilde{E}}

\newcommand{\la}{\raisebox{-.5ex}{$\stackrel{<}{\sim}$}}
\newcommand{\eq}[2]{\begin{equation}\label{#1} #2 \end{equation}}

\newcommand{\eqar}[2]{\begin{eqnarray}\label{#1} #2 \end{eqnarray}}

\begin{document}

\doublespace

\noindent

\vspace{1cm}

\begin{center}  \begin{bf}
\begin{large}
{Numerical studies of tunneling in a nonharmonic time-dependent potential}
\end{large}

\vspace{0.3in}
J. A. St\o vneng and A.-P. Jauho$^*$
\end{bf}

NORDITA

Blegdamsvej 17, DK-2100 Copenhagen \O, Denmark

\vspace{0.4in}

\begin{bf}
Abstract
\end{bf}
\end{center}

Tunneling through time-dependent potentials
is of relevance to a number of physical problems. Using a WKB analysis
Azbel' has recently studied the effects of a nonharmonic time-dependent
perturbation embedded in an opaque potential barrier. He suggests the
existence of three different regimes for transmission in such systems:
direct tunneling, activation assisted tunneling, and elevator resonant
activation. We address the same problem with a numerical technique based on
wave-packet simulations. Our numerical results are in qualitative
agreement with Azbel's picture.
The outcome depends on the characteristic time $T$ of the nonharmonic
potential. There is a transition from direct tunneling to the activation regime
around a "crossover" value $T_c$ which is determined by the
B\"{u}ttiker-Landauer time, the distance between the initial energy
and the lowest resonant level, and the amplitude of the time-dependent
perturbation. The total transmission probability is strongly enhanced
when entering the activation regime.

\vfill

\newpage

Tunneling through time-modulated potential barriers has attracted considerable
interest over the last ten years
\nocite{bl82,bl85,im85,bl86,jj89,sh89,apj90,hdr90,finn91}
\cite{bl82}-\cite{finn91}.
As argued in Refs.~\cite{az1,az2} it may be important in a
variety of physical problems, such as tunneling chemical reactions, charge
exchange
between impurity centers and resonant tunneling in semiconductors.
Until recently mostly {\em harmonic} time-dependent potentials have been
studied
theoretically. In that case the tunneling particle with initial energy $\eo$
may
absorb or emit energy quanta corresponding to the modulation frequency
$\omega$,
and the transmitted and reflected beams develop "sidebands" at energy
$\eo\pm\hbar\omega$, $\eo\pm2\hbar\omega$, \dots. Expressions for the sideband
amplitudes have been found analytically for simple rectangular barriers
\cite{bl85}, and formal generalizations to barriers of arbitrary shape were
derived
in Ref.~\cite{sh89}. In addition, numerical calculations involving wave packets
have confirmed the analytic results and completed our understanding of the
physics
involved \cite{jj89,apj90,hdr90,finn91}.

In two recent papers \cite{az1,az2} Azbel' showed that tunneling and activation
in
a {\em nonharmonic} time-dependent potential can reveal new and interesting
physics. Azbel' considered an opaque static barrier $V(x)$ with an adiabatic
time-dependent component $V_T(t)$ localized to a narrow region
in the interior of the static part \cite{op-ad}, see Fig. 1.
The potential $V_T(t)$ varies on a time scale $T$ which limits the range
of activation energies to roughly $\hbar/T$. If the potential varies
very slowly, there is virtually no activation, and the transmission probability
is
essentially unchanged from its stationary value. However, by decreasing $T$
there
may, at a given instant of time $t_0$,
be sufficient energy available to activate the particle to the
lowest instantaneous resonant level $E_r(t_0)$ in the well created by the
time-dependent potential. Since the resonant-state lifetime $\td$ is long for
opaque barriers, the activated particle will be trapped
in the resonant level and only escape after it has been lifted
above the static-barrier edge by the time-dependent
potential. The resonant level acts like an elevator for the tunneling particle;
hence the name "elevator resonant activation" (ERA) \cite{az1}. According to
Azbel'
the transition from ordinary stationary tunneling to ERA happens when
$T\sim\tau_{BL}$, where $\tau_{BL}$ is the B\"{u}ttiker-Landauer time
\cite{bl82}
for tunneling out of the well with energy $\eo$.
With an opaque barrier, as in Fig. 1, one has $\tau_{BL}\simeq \int_0^{x_0}
dx\sqrt{m^*/2[V(x)-\eo]}$. Here $V(x)$ is the static potential barrier,
$x_0$ is the classical turning point of the potential, i.e., $V(x_0)=\eo$, and
$m^*$ is the effective mass of the tunneling particle. For a rectangular
barrier
with $V(x)=V_0$ and $x_0=d$ one has $\tau_{BL}\simeq d\sqrt{m^*/2(V_0-\eo)}$.
Azbel' also predicts the existence of an intermediate
regime, "activation assisted tunneling" (AAT), where the particle is activated
to
an energy $\eat < E_r(t_0)$ before it tunnels out. Since all transmission
processes
require
tunneling into the well at energy $\eo$, the transmission probability is always
exponentially small. However, the total transmission probability is predicted
to be
greatly enhanced in the transition from direct tunneling to AAT and ERA.

In the present work we carry out numerical simulations for tunneling systems
with
parameter values appropriate to semiconductor heterostructures.
Such systems, with their large flexibility and tunability, seem to be a
promising candidate for observing the new effects predicted by Azbel'.
We study the tunneling dynamics by solving the time-dependent Schr\"{o}dinger
equation numerically \cite{numref}, and we restrict ourselves to one spatial
dimension. The initial state $\psi(x;t=0)$ is a quasi-monoenergetic minimum
uncertainty Gaussian wave packet with mean energy $\eo$, energy width
$\Delta E$, and mean velocity $v_i$. Since we have semiconductor
heterostructures in mind, suitable units of measure will be
\AA ngstr\"{o}m [\AA] for length, millielectronvolts [meV] for energy, and
femtoseconds [fs] for time. To be specific, we shall assume GaAs contacts and
AlGaAs barriers with height 230 meV. We choose the spatial width of the wave
packet
to be $\Delta x=1000$ \AA. Then
$\Delta E\simeq(\partial E/\partial k) \Delta k = \sqrt{2\eo/m^*} \hbar/ \Delta
x
\simeq 5$ meV if $\eo$ equals half the barrier height, and $m^*=0.067m_0$
(the effective mass in GaAs). Thus the initial wave packet has a
well-defined energy in all the numerical examples below, in the sense that
$\Delta E (\partial P_{tr}/\partial E)/P_{tr} << 1$ \cite{la89}.
Here $P_{tr}$ is the instantaneous transmission probability of the potential
barrier at energy $\eo$.
The initial state has its center of mass $<x(0)>$ more than 4000 \AA\ away
from the barrier structure. Thus the overlap between $\psi(x;t=0)$
and the barriers is negligible.

Our choice of initial state is not of purely academic interest.
By using modulation-doped semiconductor heterostructures
it is possible to fabricate tunneling barriers which transmit a
fairly monoenergetic beam of electrons. These electrons
can move ballistically over tens of nanometers \cite{levi} and
should be well described by wave packets of the kind used here.

The static part of the potential $V(x)$ can be made by standard
epitaxial-growth
techniques. Alternatively, the transport could take place in a two-dimensional
electron gas. In that case the potential-barrier structure is created
by applying external gate potentials.
Possible ways of producing the perturbation $V_T(t)$ could be by
means of ultra-short laser pulses or time-dependent gate potentials
\cite{kouw91}. It may be difficult to control $V_T(t)$ on the length and time
scales which are typical for the systems that we have in mind (of the order of
100 \AA\ and 100 fs, respectively), although laser pulses with a duration less
than 100 fs have already been demonstrated in several experiments.

We will present numerical results for two different kinds of potential-barrier
structures, see Fig. 2. Structure A is a single rectangular barrier of width
$W_0$
and height $V_0$. The time-dependent part of the potential acts on a segment of
width $W_1$ located at the center of the static barrier. In this region the
total potential is $V_0$ when $|t|\to\infty$ and $V_0-V_1$ for $t=t_0$.
We have used $V_1=V_0$ in all the examples below.
Within the adiabatic picture the lowest instantaneous
resonant level starts at $\erinf\equiv E_r(|t|\to\infty)$
( i.e., the "virtual" state above the barrier), falls down to
$\ero\equiv E_r(t_0)$ at $t=t_0$,
and rises back to $\erinf$ as $t\to\infty$. Structure A will be used to
investigate ERA. Structure B is similar to A, but has in addition a wide and
low barrier of height $\tilde{V}$ on each side, leading to a total width
$\tilde{W}$ for the structure. There is still a resonance in the
time-dependent well,
moving from $\erinf$ to $\ero$ and back again. The positions of $\ero$ and
$\erinf$ can be tuned by changing the well width $W_1$ (cf. Figs. 2B1 and 2B2).
In addition there are one or more
"quasi-resonant" levels $\eat_j > \tilde{V}$ $(j=1,2,...)$ corresponding to
resonances above a single rectangular barrier of width $\tilde{W}$ and height
$\tilde{V}$. The positions of these levels are quite insensitive to both the
instantaneous value and the width
of the time-dependent potential. With structure B we will
demonstrate AAT and ERA, where the outcome will
depend on the relative positions of $\ero$ and the
various $\eat_j$. In order to observe a transition from direct tunneling
to the activation regime, the initial-state energy $\eo$ must be chosen to lie
below all the resonant levels. In the opposite case, with $\ero\leq\eo$, the
elevator effect will be present for any finite $T$ since no activation energy
is required to trap the particle in the resonance \cite{az2}.
In the tables in Fig. 2 we have collected
numerical values for the most important parameters in the examples described
below. We have also plotted the transmission probability of the instantaneous
potential, both at
$t=t_0$ and $t\to\infty$. Notice the peaks in $P_{tr}$ for $E=\eat_j$
in Fig. 2B.  Although the amplitudes of these peaks depend strongly on the
value of $V_T(t)$, their positions remain more or less unchanged.
The parameter $t_0$ is chosen such that the center of mass at $t=t_0$,
$<x(t_0)>=<x(0)>+v_i t_0$, of a {\em freely
moving} wave packet with group velocity $v_i$ would coincide with the
center position of the time-dependent part of the potential.
In this way our simulations are as close as possible to the stationary
case $(\Delta x\to\infty)$ studied in Refs.~\cite{az1,az2}.

The exact form of the time-dependent part of the potential is not important.
However, in order to observe ERA, it is crucial to have a {\em nonharmonic}
variation in time. A nonharmonic time dependence implies, as we shall see
below, a
{\em continuous} activation probability. Under these conditions
a particle which has been
activated from $\eo$ to the lowest instantaneous resonant level
$\ero$, can continuously absorb
infinitesimal amounts of energy and follow the resonance as it rises from
$\ero$
to its maximum value $\erinf$. In contrast, in a purely harmonic potential with
frequency $\omega$ a particle can only absorb energy in multiples of
$\hbar\omega$,
and hence cannot be trapped in the continuously moving resonant level.
In the numerical calculations below we will use the time-dependent potential
\eq{eq1}
{V_T(t)=-\frac{V_1}{\cosh((t-t_0)/T)},}
which was also used in an example in Ref.~\cite{az1}.

Let us try to gain some insight into when the transition from direct tunneling
to
activation takes place. This must happen when the characteristic time $T$ of
the
nonharmonic potential is such that the probability $P_{act}(\eo\to\ero;T)$
for activation from $\eo$ to the lowest
resonant level $\ero$ is approximately equal to the probability
$P_{dir}(\eo)$ for direct tunneling out of the time-dependent region
with the initial energy $\eo$. An estimate of the activation
probability can be made by using standard time-dependent perturbation theory
\cite{merz}. We find
\eqar{eq2}
{P_{act}(\eo\to\ero;T)&=&|-\frac{i}{\hbar} \int_{-\infty}^{\infty} dt
<0|V_T(t)|i> e^{i(\ero-\eo)t/\hbar} |^2 \nonumber \\
&\simeq& \left[ \frac{\pi V_1 T}{\hbar\cosh[\frac{1}{2}\pi T(\ero-\eo)/\hbar]}
\right]^2,}
where $|i>$ and $|0>$ denote the initial state and the lowest resonant state,
respectively, and we have used $V_T(t)$ from Eq. (1). We have also taken into
account
that $P_{act}$ is the activation probability
{\em conditional} on the particle having tunneled into the
time-dependent region, and we assume that
the lowest resonant state can be approximated by a symmetric and
normalized wave function.
A "crossover" time $T_c$ can now be defined via
\eq{eq3}
{P_{act}(\eo\to\ero;T_c)\simeq P_{dir}(\eo).}
Since the barrier is opaque when seen from the time-dependent region, we have
\linebreak
$P_{dir}(\eo)\simeq \exp[-2d\sqrt{2m^*(V_0-\eo)}/\hbar]$, and hence
\eq{eq4}
{\left[ \frac{\pi V_1 T_c}{\hbar\cosh[\frac{1}{2}\pi
T_c(\ero-\eo)/\hbar]}\right]^2
\simeq \exp[-2d\sqrt{2m^*(V_0-\eo)}/\hbar].}
We have found that the argument of the cosh-function in Eq. (4) is typically
much larger
than unity. In that case we may rearrange Eq. (4) and obtain
a simpler equation for the crossover time,
\eq{eq5}
{T_c\simeq\frac{4}{\pi}\tau_{BL} \frac{V_0-\eo}{\ero-\eo} +
\frac{2\hbar}{\pi(\ero-\eo)} \ln \frac{2\pi V_1 T_c}{\hbar}.}
{}From Eq. (5) it is evident that the B\"{u}ttiker-Landauer time $\tau_{BL}$
sets a
scale for the crossover from direct tunneling to activated processes. As
mentioned
earlier, this was pointed out very clearly in Ref.~\cite{az1}. However, equally
important parameters in this connection are the distance from the initial
energy to the
lowest resonance, $\ero-\eo$, and the amplitude of the potential modulation,
$V_1$.

In order to analyze the tunneling and activation process in energy space,
we perform a spatial Fourier transform of the wave packet in
the asymptotic limit $t\to\infty$, thus obtaining the momentum (or energy)
distribution of the transmitted $(p>0)$ and reflected $(p<0)$ parts of the wave
packet \cite{asymtime}.

In our first example we use the structure from Fig. 2A. In Fig. 3 we have
plotted on a logarithmic scale the energy distribution of both the
transmitted (tr) and reflected (ref)
parts of the wave packet for a fairly slow potential variation ($T=182$ fs).
One can see that most of the wave packet is reflected at energy $\eo$ without
ever reaching the time-dependent region. In other words, we are in the opaque
barrier regime, consistent with the assumptions in Refs.~\cite{az1,az2}.
Next, we see that for this slowly varying potential
the energy distribution of the {\em transmitted} part of the wave packet is
also centered at $\eo$, indicating that hardly any activation has taken place.
Notice, however, the small peak (or "shoulder") at $\erinf$, i.e., at
the position of the resonant level as $t\to\infty$. These energy components are
activated to $\ero$ by frequencies $\omega\sim(\ero-\eo)/\hbar>>1/T$ in the
exponential tail of the Fourier spectrum of $V_T(t)$ and subsequently follow
the resonance adiabatically from $\ero$ to $\erinf$.
Finally, notice the symmetry of activated energy components in the
reflected and transmitted parts of the wave packet. This is seen in all our
numerical examples because activated components of the wave packet
"see" a spatially symmetric potential. Hence they have equal probability of
tunneling out in either direction and thus contribute with the same amount to
reflection and transmission.

In the following we present numerical results which illustrate the
transition from direct tunneling to activation. We will focus on the
{\em transmitted} part of the wave packet, and in Fig. 4 we have plotted
its energy distribution (now on a linear scale) for selected values of $T$,
and for the three different structures presented in Fig. 2.

The first panel in Fig. 4A is a linear plot of the transmitted distribution
in Fig. 3, and it is clear that direct tunneling dominates completely for
such a slowly varying potential. Thus the total transmission probability
for the wave packet $P_{wp}(T=182$ fs)
is very little different from the stationary value $P_{wp}(T\to\infty)$
(see the rightmost panel in Fig. 4A; solid dot and dotted line,
respectively) \cite{infprob}.  A faster potential, $T=22$ fs, yields a
strikingly different picture. The energy spectrum of the transmitted
packet is now dominated by components around $\erinf$. In this case ERA
dominates, and there is indeed a huge increase in the transmission
probability. Finally, we have included the case with $T=91$ fs, which
represents an intermediate situation with roughly equal amounts of direct
tunneling and ERA. This is in good agreement with Eq. (5), which
predicts a crossover between direct tunneling and ERA at $T_c\simeq 80$ fs.

As we decrease $T$ below 20 fs, the total transmission probability
goes through a maximum value and then falls off rapidly.
The reason is that the time-dependent potential, and hence the lowest
resonance, is sufficiently low to permit "effective"
activation only in a period of time of the order of $T$. In other words, for
decreasing $T$ a greater part of the wave packet is directly reflected
\cite{pwpmax}. By varying the initial energy $\eo$ we have
further verified that the onset of ERA scales as
expected from Eq. (5). One should also
bear in mind that when $T\la 10$ fs, we are no longer in the adiabatic regime
$\hbar/T<<\eo,V_0-\eo$, and thus outside the limits of the analytical work in
Refs.~\cite{az1,az2}.

Activation assisted tunneling, AAT, implies activation to an energy {\em below}
$\ero$ before tunneling out of the time-dependent region. In our numerical
calculations this should show up as a significant weight
between $\eo$ and $\ero$ in the energy spectrum of
the transmitted wave packet. With a structure of type A (and with an activation
probability as above) we have seen no indication of AAT. This is not
surprising since both the (static) transmission coefficient and the activation
probability behave monotonically as functions of energy between $\eo$ and
$\ero$.
Thus there is no energy between $\eo$ and $\ero$ at which the particle is
preferably transmitted, and one has a transition directly from direct
tunneling to ERA when $T$ is decreased. We should point out that this was
predicted in Ref.~\cite{az1} for a similar time-dependent potential, but with
an inverted parabola static part instead of the rectangular shape used here.

In order to observe AAT we will in the next examples use a structure of type B
(see Fig. 2B). As discussed earlier such a potential has quasi-resonances at
$\eat_j>\tilde{V}$ to which the tunneling particle may be activated. Since
the quasi-resonances $\eat_j$ do not move with the time-dependent potential,
the activated particle will eventually tunnel out with energy $\eat_j$,
and one has AAT rather than ERA. Again we use a symmetric potential, with
parameters given in the tables of Fig. 2B. The time dependence is,
as above, given by Eq. (1). In practice the low (and wide) barriers
could be made from Al$_x$Ga$_{1-x}$As, with a lower Al content $x$
than in the high (and narrow)
ones. By varying the width $W_1$ of the time-dependent region, the lowest
time-dependent resonance $\ero$ is tuned to different positions relative to the
quasi-resonances $\eat_j$. In Fig. 4B1 we use a narrow well
($W_1=18$ \AA). As a result $\ero$ lies above $\eat_1$ and $\eat_2$,
but below $\eat_3$. In Fig. 4B2 the well is somewhat wider
($W_1=35$ \AA), and $\ero$ lies below all $\eat_j$.
We have again plotted the energy distribution of the transmitted part of the
wave
packet for various values of $T$. In the case of the narrow well the direct
tunneling component is significant for all $T$. This shows up as a
peak at $E\simeq\eo$ in the transmitted energy distribution.
Furthermore the selected examples in Fig. 4B1
illustrate AAT to $\eat_1$ when $T=91$ fs, AAT to $\eat_1$ and $\eat_2$
(the latter only as a weak shoulder in the energy distribution) when
$T=46$ fs, and a combination of AAT to $\eat_1$ and $\eat_2$ and ERA from
$\ero$ to $\eat_3$ and above when $T=6$ fs. With the wider well we have
observed substantial
activation to $\ero$ followed by ERA already for $T\sim 350$ fs.
When $T=182$ fs (first panel, Fig. 4B2), most of
the activated components follow the resonance to $\eat_1$ and tunnel out.
Reducing $T$ to 46 fs results in ERA to $\eat_1$ and $\eat_2$. In that case the
lifetime of the lowest quasi-resonance, $\tilde{\tau}_1\simeq 70$ fs, is so
long
that many of the "elevating" wave-packet components
are trapped in the moving resonance and do not
tunnel out at $\eat_1$. However, the second quasi-resonance has a lifetime
$\tilde{\tau}_2\simeq 35$ fs which is shorter than $T$. Thus the elevating
components that pass through the level at $\eat_1$
tend to tunnel out at $\eat_2$. Upon further reduction, to $T=6$ fs, we
observe ERA to $\eat_1$, $\eat_2$ and $\eat_3$, and some components even
follow the resonance all the way up to $\erinf$ before they tunnel
out. In Figs. 4B1 and 4B2 (rightmost panel) we have also
plotted the total transmission probability as function of
$T$ for the two cases discussed above. As in Fig. 4A there is a large
enhancement
of $P_{wp}$ when $T$ is made short enough to allow for AAT or ERA, as predicted
in Refs.~\cite{az1,az2}. The crossover values calculated from Eq. (5)
are $T_c= 137$ fs for Fig. 4B1 and
$T_c=285$ fs for Fig. 4B2, both in good agreement with what is observed.
For sufficiently small values of $T$ the transmission
probability goes through a maximum and falls off rapidly,
for the same reason as discussed in connection with Fig. 4A.

We have also examined other time-dependent
potentials, both harmonic and nonharmonic ones. As expected the qualitative
features
in the examples above stay unchanged as long as the activation probability is a
continuous function of energy.
On the other hand, with a purely harmonic potential, $V_1\cos\omega t$,
there is no elevator effect but simply activation and tunneling at energies
$\eo+n\hbar\omega$ $(n=0,\pm1,\pm2,\dots)$.

In conclusion, we have studied tunneling through a nonharmonic time-dependent
potential by letting Gaussian wave packets scatter off the barrier. Analysis of
the energy distribution of the transmitted wave packets for suitably chosen
tunneling structures confirms Azbel's prediction
of a crossover from direct tunneling to elevator resonant activation and
activation
assisted tunneling. An estimate shows that activation dominates over direct
tunneling when the characteristic time of the time-dependent potential is
reduced below a crossover value which involves the B\"{u}ttiker-Landauer time,
the
distance from initial energy to the lowest resonant level, and the amplitude of
the
time-dependent potential. The total transmission probability is strongly
enhanced, in our numerical calculations by one or two
orders of magnitude, when comparing activation with direct tunneling.

We would like to acknowledge K. Flensberg for useful discussions.

\vspace{3cm}
\noindent
$^*$ Also at Microelectronics Centre, Technical University of Denmark,
2800 Lyngby, Denmark.

\vspace{3cm}
\begin{Large}
\begin{bf}
\noindent
Figure Captions
\end{bf}
\end{Large}

\vspace{.5cm}
\noindent
FIG.1. One-dimensional potential barrier $V(x)$ with a time-dependent component
$V_T(t)$ embedded near the center. $x_0$ denotes the classical turning point
for incoming energy $\eo$. Solid and dotted vertical lines
in the well illustrate three possible processes, as
described in the main text: direct tunneling at $\eo$ (DT), activation to
$\eat$
and tunneling (AAT), activation to $E_r(t_0)$ and elevation to
$E_r(\infty)$ (ERA).

\vspace{.5cm}
\noindent
FIG.2. Potential-barrier structures, static transmission probabilities versus
energy $P_{tr}(E)$, and tables with potential-barrier parameters and
resonant-level
energies for the
two types of structure used in our numerical calculations. (A) A
rectangular static barrier of height $V_0$ and width $W_0$. The time-dependent
part
is located at the center of the static barrier and has width
$W_1$ and maximum value $-V_1$. We have used $V_1=V_0$ in all numerical
calculations.
The lowest instantaneous resonance moves between $\erinf$ and
$\ero$. (B) Similar to the structure in A, but here we have in
addition wide and low barriers of
height $\tilde{V}$ on each side. The total width is $\tilde{W}$. In addition to
the moving resonance there are "quasi-resonances" at $\eat_1,\eat_2,\dots$.
Figs. B1 and B2 represent results for structures which are identical apart from
having different well width $W_1$. The levels in the well labeled with
1 and 2 denote the extremes $\ero$ and $\erinf$ of the moving resonance
(for the narrow and wide well, respectively).
The transmission probability is in all three cases plotted for
the maximum and minimum value of the
time-dependent potential: Solid lines are for $t\to\infty$
whereas dashed lines are for $t=t_0$, i.e., for $V=V_0$ and $V=0$,
respectively, in the time-dependent region.
The peaks in $P_{tr}$ are associated with the various resonant levels
$\ero,\erinf,\eat_1,\eat_2,\dots$.

\vspace{.5cm}
\noindent
FIG.3. Logarithm of the energy distribution $|\Phi(E)|^2$ of the
reflected (ref) and transmitted (tr) parts of the wave packet after
scattering off the structure in Fig. 2A. (Note that the $E$-axis has
positive values in both directions.)
The characteristic time of the
time-dependent potential is $T=182$ fs. Both parts of the wave packet
are dominated by energy components around the initial energy
$\eo\simeq81$ meV. The "shoulders" with their maxima close to
$\erinf\simeq248$ meV represent energy components that have been
activated to the lowest instantaneous resonance $\ero\simeq127$ meV
and elevated from $\ero$ to $\erinf$ (ERA).

\vspace{.5cm}
\noindent
FIG.4. Numerical results corresponding to the structures presented in Fig.
2. Reading horizontally, the first three panels show the energy distribution
(on a linear scale)
of the transmitted part of the wave packet for selected values of $T$.
Note that the scale along the vertical axis varies from curve to curve.
The rightmost panel shows the total transmission probability as function of
$T$.
These curves result from integrating the transmitted energy distribution
for each value of $T$. The solid dots represent the selected values of $T$.
The dotted horizontal line (for the case B2 this line lies very close
to $P_{wp}=0$) represents the {\em stationary}
value of the transmission probability, $P_{wp}(T\to\infty)$,
for the particular structure.
(A) Results for the rectangular barrier in Fig. 2A.
The first frame ($T=182$ fs) is a linear plot of the transmitted part
in Fig. 3. (B) Results for the structures of Fig. 2B,
for the narrow well (B1) and the wide well (B2).
In all cases the peaks in the transmitted energy distribution are
connected to the initial energy $\eo$ or to the various
resonant levels, which can be identified from the peaks in the
static transmission probabilities in Fig. 2.


\begin{thebibliography}{99}
\bibitem{asymtime} In practice we must choose an
"observation time" $t_{obs} >> t_0$. However, $t_{obs}$ should be
small enough to avoid reflections from the endpoints of
the finite spatial line.
\bibitem{infprob}In fact, $P_{wp}(T\to\infty)>P_{wp}(T=182$ fs). This is as
expected since $T\to\infty$ represents the stationary situation with $V=0$
in the otherwise time-dependent region. As a result the effective tunneling
barrier is lower for $T\to\infty$ than for any finite $T$, and $P_{wp}(\infty)$
exceeds $P_{wp}(T)$ for those $T$ where the amount of activation
is negligible.
\bibitem{pwpmax} This effect seems to be contained in Eq. (2) which
indicates a maximum in the activation probability $P_{act}(T)$ for a
finite $T_m$. However, Eq. (2) is based on perturbation theory and is
only valid if $P_{act}(T)<<1$. For typical values of $V_1$ and
$\ero-\eo$, $T_m$ is well below the range of validity of Eq. (2) and
cannot be used as a quantitative estimate of the maximum position in
$P_{wp}(T)$.
\end{thebibliography}
\end{document}